# Monte Carlo method for polarized radiative transfer in gradient-index media

J. M. Zhao[a], J. Y. Tan[b], L. H. Liu[a,b]*

[a] *School of Energy Science and Engineering, Harbin Institute of Technology, 92 West Dazhi Street, Harbin 150001,*

*People's Republic of China*

[b] *School of Automobile Engineering, Harbin Institute of Technology at Weihai, 2 West Wenhua Road, Weihai*

*264209, People's Republic of China*

**Abstract**

Light transfer in gradient-index media generally follows curved ray trajectories, which will cause light beam to converge or diverge during transfer and induce the rotation of polarization ellipse even when the medium is transparent. Furthermore, the combined process of scattering and transfer along curved ray path makes the problem more complex. In this paper, a Monte Carlo method is presented to simulate polarized radiative transfer in gradient-index media that only support planar ray trajectories. The ray equation is solved to the second order to address the effect induced by curved ray trajectories. Three types of test cases are presented to verify the performance of the method, which include transparent medium, Mie scattering medium with assumed gradient index distribution, and Rayleigh scattering with realistic atmosphere refractive index profile. It is demonstrated that the atmospheric refraction has significant effect for long distance polarized light transfer.

______________________________________________________________

*Corresponding author. Tel.: +86-451-86402237; fax: +86-451-86221048.

*E-mail addresses:* jmzhao@hit.edu.cn (J.M. Zhao), tanjy@hit.edu.cn (J.Y. Tan), lhliu@hit.edu.cn (L. H. Liu).

## 1. Introduction

The refractive index of a material is dependent on its component concentration, temperature, pressure, density, etc. When the dependent properties are distributed inhomogeneously, the refractive index of the media will become a function of space. Examples of participating media with gradient refractive index distribution include the natural media such as earth's (or other planets') atmosphere, the ocean water, the hot air/gas of a flame, and artificial materials, such as gradient index lens, gradient index optical fiber, etc. Theory of radiative transfer in gradient-index media is fundamental to study the light transport in such media, and is of significant importance for the applications (or potential applications) in the fields of atmospheric remote sensing [1-4], atmospheric optics [5-9], and non-contact measurement of temperature distribution in flames [10-12], etc.

In gradient-index media, light beam will generally travel along curved trajectories determined by Fermat's principle [13-15]. The gradient index distribution will induce special effects in radiative transfer. The variation of refractive index along the ray trajectory will cause light beam to converge or diverge and hence influence the transport of radiative intensity. Furthermore, the curved ray trajectory will induce rotation of polarization ellipse as described by the Rytov's field-vector rotation law [14]. Hence polarized radiative transfer in gradient-index media is much more complicated than in uniform-index media. The vector radiative transfer theory developed by Chandrasekhar [16] is the basis for analyzing polarized radiative transfer in media with uniform refractive index distribution. However, it is not appropriate for gradient-index media, since the aforementioned effects induced by the curved ray trajectory are not considered in the formulation. Up till now, there were very few works on polarized radiative transfer in gradient-index media. To our knowledge, Lau and Watson [17] were the first to present a derivation of the radiative transfer equation for polarized light transport in gradient-index media. Recently, Zhao et. al. [18] presented a new derivation and a more generalized form of the equation of polarized radiative transfer in gradient-index media was given.

Solution methods for polarized radiative transfer in uniform-index media have been studied for a long time, and many successful methods have been developed and used widely, such as Monte Carlo method [19-23], vector discrete ordinates method [24-26], adding-doubling method [27], etc. In recent years, scalar radiative transfer in gradient-index media has attracted the interest of many researchers, and many simulation methods were developed, such as curved raytracing method [28-30], Monte Carlo method [31, 32], finite volume method [33-35], finite element method [36-38], DRESOR method [39], to name a few. However, there are still very few works on the solution of polarized radiative transfer in gradient-index media, though there were some works on the polarized radiative transfer in multilayer medium, such as, the work of Garcia [40, 41]. Most recently, Ben et al. [42] presented a method to simulate polarized radiative transfer in gradient-index media, in which a multilayer slab model was used to approximate the gradient index distribution and then the traditional vector radiative transfer equation is solved by a MC method in each layer. This approximation is inappropriate when a light beam is launched in a direction parallel to the layer. As shown in Fig. 1, the actual ray path in gradient-index media should bend toward the gradient direction indicated by the ray equation [43] as illustrated in Fig. 1(a), whereas the multilayer model predicts a straight path perpendicular to the gradient direction as illustrated in Fig. 1(b).

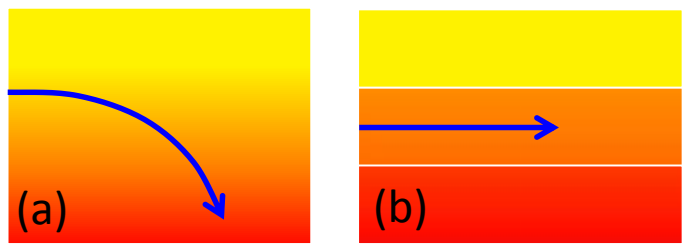

**FIG. 1.** Illustration of ray path when a light beam transfers **(a)** in gradient-index medium, and **(b)** in a multilayer medium.

In this work, a Monte Carlo (MC) method is presented to simulate polarized radiative transfer in gradient-index media. The ray equation is solved to the second order to address the effect induced by curved ray path. The gradient-index media that support planar ray trajectories are considered for demonstration. Three types of test cases are presented to verify the performance of the presented MC method and for comparison with other models, which include a case with analytical solution, a case with Mie scattering and an assumed gradient index distribution, and a case concerning Rayleigh scattering with realistic atmosphere refractive index profile.

## 2. Equation of polarized radiative transfer in gradient-index media

A light beam with arbitrary polarization can be described by the Stokes parameters vector $\mathbf{I} = (I, Q, U, V)^T$ [16, 44, 45]. Additional geometric effect exists when light transfer in gradient-index medium, the curved ray paths will cause the Stokes parameters changing during transport even when the medium is transparent. Generally, the governing equation of polarized radiative transfer in an absorbing, emitting and scattering gradient-index medium (GVRTE) can be written along the ray trajectory as [18]

$$n^2 \frac{\mathrm{d}}{\mathrm{d}s}\left[\frac{\mathbf{I}}{n^2}\right] + \left(\boldsymbol{\kappa}_e + \mathbf{R}\right)\mathbf{I} = \boldsymbol{\kappa}_a \mathbf{I}_b + \frac{1}{4\pi}\int_{4\pi} \mathbf{Z}(\boldsymbol{\Omega}', \boldsymbol{\Omega})\mathbf{I}(s, \boldsymbol{\Omega}')\mathrm{d}\Omega \tag{1}$$

where $s$ is the coordinate along the ray trajectory, $n$ is the refractive index, $\boldsymbol{\kappa}_e$ is the extinction matrix, $\boldsymbol{\kappa}_a$ is the emission matrix, $\mathbf{I}_b$ is the Stokes parameters vector of emission in the refractive medium, $\mathbf{Z}$ is the scattering phase matrix and $\mathbf{R}$ is a matrix accounting for the rotation of the polarization ellipse determined by the Rytov's law [14] defined as

$$\mathbf{R} = \begin{bmatrix} 0 & 0 & 0 & 0 \\ 0 & 0 & 2\upsilon & 0 \\ 0 & -2\upsilon & 0 & 0 \\ 0 & 0 & 0 & 0 \end{bmatrix} \tag{2}$$

in which $\upsilon$ stands for the torsion of the trajectory.

If the gradient-index medium only supports ray trajectories being planar curves, then $\upsilon = 0$ and the GVRTE can be simplified as

$$n^2 \frac{\mathrm{d}}{\mathrm{d}s}\left[\frac{\mathbf{I}}{n^2}\right] + \boldsymbol{\kappa}_e \mathbf{I} = \boldsymbol{\kappa}_a \mathbf{I}_b + \frac{1}{4\pi}\int_{4\pi} \mathbf{Z}(\boldsymbol{\Omega}',\boldsymbol{\Omega})\mathbf{I}(s,\boldsymbol{\Omega}')\mathrm{d}\Omega \qquad (3)$$

There are two groups of media possess this property, such as the case that the refractive index distribution satisfies $\mathbf{t}_0 \times \nabla n = 0$, where $\mathbf{t}_0$ is a fixed vector and the case of radial gradient index distribution with spherical symmetry. For isotropic random media, the apparent extinction and absorption process should be polarization independent, then Eq. (3) can be further written as

$$n^2 \frac{\mathrm{d}}{\mathrm{d}s}\left[\frac{\mathbf{I}}{n^2}\right] + \beta\mathbf{I} = \kappa_a \mathbf{I}_b + \beta\frac{\omega}{4\pi}\int_{4\pi} \bar{\mathbf{Z}}(\boldsymbol{\Omega}',\boldsymbol{\Omega})\mathbf{I}(s,\boldsymbol{\Omega}')\mathrm{d}\Omega \qquad (4)$$

where $\beta$, $\kappa_a$ and $\omega$ are the extinction coefficient, absorption coefficient and single scattering albedo, respectively, and $\bar{\mathbf{Z}}(\boldsymbol{\Omega}',\boldsymbol{\Omega})$ is the normalized scattering phase matrix satisfying

$$\int_{4\pi} \bar{Z}_{11}(\boldsymbol{\Omega}',\boldsymbol{\Omega})\mathrm{d}\Omega = 4\pi \qquad (5)$$

Note that $\bar{Z}_{11}$ is also the corresponding phase function for scalar radiative transfer.

In this study, only the gradient-index media that support planar ray trajectories are considered. A curved ray tracing MC method is presented to simulate the transport of polarized photons in gradient-index medium described by Eq. (4).

## 3. Monte Carlo simulation

The Stokes parameters for quasi-monochromatic beam are given as [46]

$$\mathbf{I} = \begin{bmatrix} I \\ Q \\ U \\ V \end{bmatrix} = \begin{bmatrix} \langle E_l E_l^* + E_r E_r^* \rangle \\ \langle E_l E_l^* - E_r E_r^* \rangle \\ \langle E_l E_r^* + E_r E_l^* \rangle \\ i\langle E_l E_r^* - E_r E_l^* \rangle \end{bmatrix} \qquad (6)$$

where $E_l$ and $E_r$ denotes parallel and perpendicular components of electric field vector, respectively, $i$ is the imaginary unit, and the angular brackets denotes time average over a long time compared to the period.

Here we choose the parallel direction vector $\mathbf{p}_\parallel$, perpendicular direction vector $\mathbf{p}_\perp$ and the beam propagation direction vector $\mathbf{\Omega} = \mathbf{p}_\perp \times \mathbf{p}_\parallel$ to form a left-hand reference frame as was used in Refs. [46, 47]. The local direction vectors $\mathbf{p}_\parallel$ and $\mathbf{p}_\perp$ are defined as the direction being parallel and perpendicular to the meridian plane, which is defined based on the $z$-axis of the global coordinates system. The definitions of global coordinates system and the local direction vectors are shown in Fig. 2. The definition of the reference frame for polarization is also used for beam of arbitrary polarization.

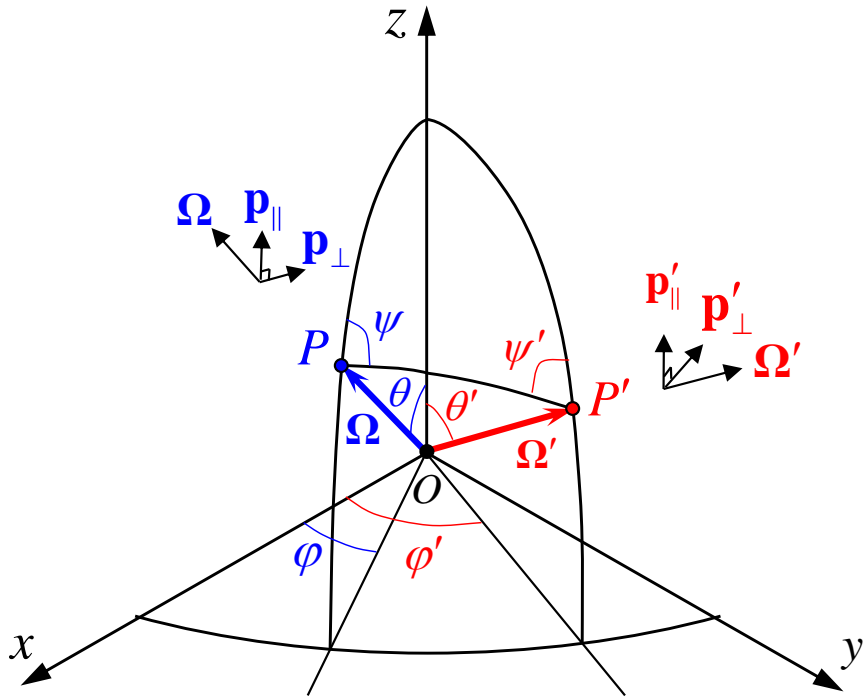

**FIG. 2**. Geometry for the definition of polarization reference frame and definition of angles in scattering calculation.

In the MC simulation, the state of a photon bundle is traced in the medium. For steady state polarized radiative transfer, a photon bundle carries both the power and the information of polarization. In the present approach, the power carried by a photon bundle is described by a weight variable $W$, and the polarization state is described by a Stokes parameters vector with $I = 1$ as $\mathbf{S} = (1, \bar{Q}, \bar{U}, \bar{V})$, called the polarization state vector. The Stokes parameters vector of a photon bundle is then obtained as $\mathbf{I} = W\,\mathbf{S}$.

Due to moving, scattering and absorption, the properties of a photon bundle, such as location,

propagation direction, polarization state and weight will change when it transfers in the gradient-index media. In the following, the MC simulation of the transport process is presented.

### 3.1. Moving in gradient-index medium

The moving of photon bundle is traced in a global Cartesian coordinate system. In gradient-index media, photon travels in curved ray path determined by the ray equation [43].

$$\frac{\mathrm{d}}{\mathrm{d}s}\left(n\boldsymbol{\Omega}\right) = \nabla n \tag{7}$$

The propagation direction vector can be calculated by definition as $\boldsymbol{\Omega} = \dfrac{\mathrm{d}\mathbf{r}}{\mathrm{d}s}$, where $\mathbf{r}$ is the coordinates vector in a global Cartesian coordinate system. The key issue to move the photon bundle is to obtain the new global Cartesian coordinates $\mathbf{r}$ after a moving of short distance $\Delta s$ along the ray trajectory. Using Taylor expansion along ray coordinates at initial location $s_0$, the new location $\mathbf{r}(s)$ in Cartesian coordinate system after a small propagation distance $\Delta s = s - s_0$ can be expressed as

$$\mathbf{r}(s) = \mathbf{r}(s_0) + \left[\frac{\mathrm{d}\mathbf{r}(s)}{\mathrm{d}s}\right]_{s=s_0}\Delta s + \frac{1}{2}\left[\frac{\mathrm{d}^2\mathbf{r}(s)}{\mathrm{d}s^2}\right]_{s=s_0}(\Delta s)^2 + ... \tag{8}$$

To account for the curvature of ray path, Eq. (8) needs to be at least truncated to the second order. The first and second term in the expansion are readily known, where $\mathbf{r}(s_0) = \mathbf{r}_0$ and $\left[\dfrac{\mathrm{d}\mathbf{r}(s)}{\mathrm{d}s}\right]_{s=s_0} = \boldsymbol{\Omega}_0$ stands for the location and propagation direction of the photon bundle before moving, respectively. To calculate the third term, the ray equation can be written as

$$\frac{\mathrm{d}}{\mathrm{d}s}\left(n\frac{\mathrm{d}\mathbf{r}}{\mathrm{d}s}\right) = n\frac{\mathrm{d}^2\mathbf{r}}{\mathrm{d}s^2} + \frac{\mathrm{d}\mathbf{r}}{\mathrm{d}s}\frac{\mathrm{d}n}{\mathrm{d}s} = \nabla n \tag{9}$$

As such, the third term can be determined from the ray equation and calculated in Cartesian coordinates system from

$$\left[\frac{\mathrm{d}^2\mathbf{r}(s)}{\mathrm{d}s^2}\right]_{s=s_0} = \left[\nabla\ln n\right]_{\mathbf{r}=\mathbf{r}_0} - \boldsymbol{\Omega}_0\left(\boldsymbol{\Omega}_0\bullet\left[\nabla\ln n\right]_{\mathbf{r}=\mathbf{r}_0}\right) \tag{10}$$

Note the relation $\mathrm{d}/\mathrm{d}s = \mathbf{\Omega} \cdot \nabla$ is used in the derivation.

The new location of the photon bundle after moving a short distance $\Delta s$ can then be calculated based on Eq. (8) truncated to second order with the known quantities at the initial location, namely, location vector $\mathbf{r}_0$, propagation direction vector $\mathbf{\Omega}_0$ and the gradient of index $[\nabla \ln n]_{\mathbf{r}=\mathbf{r}_0}$. It is noted that $\Delta s$ cannot be taken too long in order to assure good accuracy of curved ray tracing because of the truncated series approximation.

The free traveling path length $l$ of a photon bundle can be determined in the same way as in traditional MC method for radiative transfer. Considering the extinction coefficient is homogeneous, the free traveling path length $l$ of the photon bundle can be simulated as [20, 48]

$$l = -\frac{1}{\beta} \ln R_l \qquad (11)$$

where $R_l$ is a random number of uniform distribution in $[0,1]$. After $l$ is determined, the photon bundle is then moved along the curved ray trajectory using Eq. (8) step by step until $\sum_i \Delta s_i = l$, where $\Delta s_i$ is the tracing distance of each step. In the implementation, the tracing step size is fixed except for the final step to reach the free traveling path, where a tuned step size is used to make the total tracing distance exactly equal to $l$.

### 3.2. Interaction: absorption, scattering and reflection

Three kinds of interaction are considered here, namely, absorption and scattering in media, and reflection at interfaces. As indicated by the GVRTE, the Monte Carlo models for absorption and scattering of polarized light in uniform-index media should also apply for gradient-index media. There were a few works on Monte Carlo method for polarized radiative transfer, such as Refs. [19-22]. Though the basic principle is the same, there are a few variants in the detailed implementation of Monte Carlo methods [20]. In the following, the implementation of present Monte Carlo simulation is presented.

During the interaction with media, the power of a photon bundle (represent as a weight variable $W$) will gradually be reduced due to absorption and finally be terminated when it is smaller than a given threshold. When interaction with media occurs, namely, the free traveling distance is reached, some portion of the power is absorbed and some portion is scattered. The absorbed power is calculated as $(1-\omega)W$ and the new weight of the photon bundle is set as $W' = \omega W$. The tracing of the photon bundle will be terminated if power of the photon bundle is too small.

If the weight $W'$ is greater than the threshold, the photon bundle is then scattered. It is assumed that the photon bundle is scattered to a specific direction as a whole. The scattered direction and change of polarization state are two key issues in the simulation of scattering process, which are determined from the scattering phase matrix. By definition, the relation between the scattered Stokes parameters $\mathbf{I}$ and the incident Stokes parameters vectors $\mathbf{I}_i = (I_i, Q_i, U_i, V_i)$ is

$$\mathbf{I}(\mathbf{\Omega}) = \mathbf{Z}(\mathbf{\Omega}', \mathbf{\Omega})\mathbf{I}_i(\mathbf{\Omega}') \tag{12}$$

where $\mathbf{\Omega}'$ and $\mathbf{\Omega}$ denotes the incident direction and scattering direction respectively. The phase matrix $\mathbf{Z}(\mathbf{\Omega}', \mathbf{\Omega})$ is usually expressed based on transformation of the scattering matrix $\mathbf{P}$ as [16, 44]

$$\mathbf{Z}(\mathbf{\Omega}', \mathbf{\Omega}) = \mathfrak{R}(\pi - \psi)\mathbf{P}(\mathbf{\Omega}', \mathbf{\Omega})\mathfrak{R}(-\psi') \tag{13}$$

in which the rotation matrix $\mathfrak{R}(\psi)$ is defined as

$$\mathfrak{R}(\psi) = \begin{bmatrix} 1 & 0 & 0 & 0 \\ 0 & \cos 2\psi & \sin 2\psi & 0 \\ 0 & -\sin 2\psi & \cos 2\psi & 0 \\ 0 & 0 & 0 & 1 \end{bmatrix} \tag{14}$$

The definitions of angles for scattering calculation are shown in Fig. 2.

The first step is to determine the scattered direction. It is convenient to use a new local coordinate system, with $\mathbf{p}'_{\parallel}$ as the local $x$ direction (direction vector $\hat{\mathbf{e}}_x$), $\mathbf{\Omega}'$ as the local $z$ direction (direction vector $\hat{\mathbf{e}}_z$ ), and the $y$ direction (direction vector $\hat{\mathbf{e}}_y$ ) is determined to make $x$-$y$-$z$ a right hand frame. In this local coordinate system, the scattered direction vector can thus be written as

$$\mathbf{\Omega} = \sin\hat{\theta}\cos\hat{\varphi}\hat{\mathbf{e}}_x + \sin\hat{\theta}\sin\hat{\varphi}\hat{\mathbf{e}}_y + \cos\hat{\theta}\hat{\mathbf{e}}_z \qquad (15)$$

where $\hat{\theta}$ and $\hat{\varphi}$ are the polar and azimuthal angle of $\mathbf{\Omega}$ defined based on $\hat{\mathbf{e}}_z$ and $\hat{\mathbf{e}}_x$ in this local coordinate system, and

$$\hat{\mathbf{e}}_z = \mathbf{\Omega}' , \quad \hat{\mathbf{e}}_y = -\mathbf{n}_{OZP'} , \quad \hat{\mathbf{e}}_x = \hat{\mathbf{e}}_y \times \hat{\mathbf{e}}_z = \mathbf{\Omega}' \times \mathbf{n}_{OZP'} \qquad (16)$$

in which $\mathbf{n}_{OZP'}$ denotes the normal vector of the plane $OZP'$ as shown in Fig. 2. Eq. (15) can be finally written as

$$\mathbf{\Omega} = \begin{pmatrix} \Omega'_x \cos\hat{\theta} + \dfrac{\sin\hat{\theta}}{\sqrt{1-\Omega_z'^2}}\left(\Omega'_x\Omega'_z\cos\hat{\varphi} - \Omega'_y\sin\hat{\varphi}\right) \\ \Omega'_y \cos\hat{\theta} + \dfrac{\sin\theta}{\sqrt{1-\Omega_z'^2}}\left(\Omega'_y\Omega'_z\cos\hat{\varphi} + \Omega'_x\sin\hat{\varphi}\right) \\ \Omega'_z \cos\hat{\theta} - \sin\hat{\theta}\cos\hat{\varphi}\sqrt{1-\Omega_z'^2} \end{pmatrix} \qquad (17)$$

Note that a similar expressions of $\mathbf{\Omega}$ in the local coordinates was presented in Ref. [20]. From Eq. (17), the scattered direction can be calculated if a statistical model of $\hat{\theta}$ and $\hat{\varphi}$ is built based on the scattering matrix $\mathbf{P}$ . Here the six element scattering matrix for spherical particle is considered, namely,

$$\mathbf{P}(\cos\Theta) = \begin{bmatrix} \Phi_1 & \Phi_2 & 0 & 0 \\ \Phi_2 & \Phi_5 & 0 & 0 \\ 0 & 0 & \Phi_3 & \Phi_4 \\ 0 & 0 & -\Phi_4 & \Phi_6 \end{bmatrix} \qquad (18)$$

where $\cos\Theta = \mathbf{\Omega} \bullet \mathbf{\Omega}'$ . Note that $\psi' = \hat{\varphi}$ in the local coordinate system, using the relations Eqs. (12), (13) and (18), the scattered intensity at direction $\mathbf{\Omega}(\hat{\theta},\hat{\varphi})$ can be calculated in the local coordinate system from

$$I(\mathbf{\Omega}) = \Phi_1(\cos\hat{\theta})I_i + \Phi_2(\cos\hat{\theta})\left[Q_i\cos2\hat{\varphi} - U_i\sin2\hat{\varphi}\right] \qquad (19)$$

Because the form of phase function is usually complicated, here rejection method is used to select the scattering directions that satisfy the probability distribution given by Eq. (19). A valid random scattering direction is chosen only when $R_r < I(\Omega)/I_{max}$, otherwise the direction is rejected, where $R_r$ is a uniformly distributed random number in $[0,1]$. The maximum value of the scattered intensity $I_{max}$ is chosen as

$$I_{max} = \max(\Phi_1)I_i + \max(|\Phi_2|)(|Q_i|+|U_i|) \tag{20}$$

Note that the value $\max(\Phi_1)$ and $\max(|\Phi_2|)$ can be calculated beforehand. The uniformly distributed random directions used in the rejection method are defined by the local polar angle $\hat{\theta}$ and local azimuthal angle $\hat{\varphi}$, which are generated from

$$\hat{\theta} = \arccos(1-2R_{\hat{\theta}}), \ \hat{\varphi} = 2\pi R_{\hat{\varphi}} \tag{21}$$

where $R_{\hat{\theta}}$ and $R_{\hat{\varphi}}$ are the uniformly distributed random number in $[0,1]$.

The second step is to determine the polarization state of the scattered photon bundle. Based on Eqs. (12), (13) and (19), the polarization state vector of the scattered photon bundle $\mathbf{S}$ can be calculated from $\mathbf{S} = \tilde{\mathbf{S}}/\tilde{I}$, where $\tilde{I}$ denotes the intensity component of $\tilde{\mathbf{S}}$, defined as

$$\tilde{\mathbf{S}} = \mathfrak{R}(s_g\hat{\varphi}'')\mathbf{P}(\cos\hat{\theta})\mathfrak{R}(-\hat{\varphi})\mathbf{S}_i \tag{22}$$

in which $\mathbf{S}_i$ is the polarization state vector of the incident photon, $\hat{\varphi}'' = \arccos(\mathbf{n}_{OZP}\cdot\mathbf{n}_{OPP'})$, and $s_g$ is a sign variable defined as

$$s_g = \begin{cases} 1, & \hat{\varphi} \in [0,\pi] \\ -1, & \hat{\varphi} \in (\pi,2\pi] \end{cases} \tag{23}$$

In the following, we consider the reflection at the diffuse reflective boundary and an interface between two media. When a photon bundle reaches a boundary or interface, it is considered to be reflected, transmitted or absorbed as a whole. At a Lambertian reflection boundary with reflectance $\rho$, the weight is modified as $W' = \rho W$, where $W$ is the weight just before reflection. The photon bundle is totally depolarized after the reflection, hence the polarization state vector is set as $\mathbf{S} = (1,0,0,0)$. The reflected directions are chosen randomly to satisfy Lambert's law. The polar angle $\theta$ and azimuthal angle $\varphi$ defined in the local coordinate system (by taking the normal vector as the local $z$-direction) are chosen as

$$\theta = \arccos\left(\sqrt{1 - R_\theta}\right)$$
$$\varphi = 2\pi R_\varphi$$

(24)

where $R_\theta$ and $R_\varphi$ are the uniformly distributed random number in $[0, 1]$.

When the photon bundle is incident from medium 1 with refractive index $n_1$ to medium 2 with refractive index $n_2$, the reflectivity $\rho_{fr}$ is calculated from (note this equation is obtained from the Muller matrix for reflection)

$$\rho_{fr}(\theta_i) = \frac{1}{2}\left[\left(|r_p|^2 + |r_s|^2\right) + \left(|r_p|^2 - |r_s|^2\right)\bar{Q}_i\right]$$

(25)

where $\theta_i$ denotes the incident angle, $\bar{Q}_i = Q_i / I_i$ is the second component of the incident polarization state vector $\mathbf{S}_i$, $r_p$ and $r_s$ are the amplitude reflectivity of the parallel and perpendicular polarization, respectively, and are calculated using Fresnel equations as [1]

$$r_p = \frac{m^2\cos\theta_i - \sqrt{m^2 + \cos^2\theta_i - 1}}{m^2\cos\theta_i + \sqrt{m^2 + \cos^2\theta_i - 1}}, \quad r_s = \frac{\cos\theta_i - \sqrt{m^2 + \cos^2\theta_i - 1}}{\cos\theta_i + \sqrt{m^2 + \cos^2\theta_i - 1}}$$

(26)

where $m = n_2 / n_1$. The weight of the reflected photon bundle is set as $W' = \rho_{fr}(\theta_i)W$ after reflection. A generated uniform distribution random number $R_{fr} \in [0, 1]$ is used to determine if the reflection or transmission occurs, if $R_{fr} < \rho_{fr}(\theta_i)$, the photon bundle is determined to reflect and otherwise it is transmitted. The polarization state after reflection is calculated using the reflection Muller matrix $\mathbf{M}_R(\theta_i)$ as

$$\mathbf{S} = \frac{1}{\rho_{fr}(\theta_i)}\mathbf{M}_R(\theta_i)\mathbf{S}_i$$

(27)

where $\mathbf{S}_i$ denotes the incident polarization state vector, and $\mathbf{M}_R$ is defined as

$$\mathbf{M}_R(\theta_i) = \begin{bmatrix} (|r_p|^2 + |r_s|^2)/2 & (|r_p|^2 - |r_s|^2)/2 & 0 & 0 \\ (|r_p|^2 - |r_s|^2)/2 & (|r_p|^2 + |r_s|^2)/2 & 0 & 0 \\ 0 & 0 & \mathrm{Re}(r_p r_s^*) & \mathrm{Im}(r_p r_s^*) \\ 0 & 0 & -\mathrm{Im}(r_p r_s^*) & \mathrm{Re}(r_p r_s^*) \end{bmatrix}$$

(28)

Note that a similar form of $\mathbf{M}_R$ was also presented in Ref. [49]. It is noted that Eq. (27) is used because the first component of the polarization state vector should always be unit by definition.

### *3.3. Photon detection and post-processing*

For steady state MC simulation, the weight of a photon bundle is considered to have dimension of power (watt). The amount of power received by a detector is calculated by the summation of all the weights of the photon bundles hit on the detector,

$$P_{rec} = P_{src} D_f \tag{29}$$

where $P_{src}$ denotes the power of the source, $D_f = \dfrac{1}{NW_0}\sum_k W_k$ is a statistically calculated photon-distribution-factor by using the weight of detected photon bundles, $W_k$ denotes the weight of the $k$-th detected photon bundle, $W_0$ is the initial weight of a photon bundle (assumed $W_0 = 1$ in the simulation), $N$ denotes the total number of photon bundles launched.

For a detector with area $A$, the polarized flux density $\mathbf{q}$ is obtained as

$$\mathbf{q} = \frac{P_{src}}{A}\mathbf{D}_f \tag{30}$$

where $\mathbf{D}_f = \dfrac{1}{NW_0}\sum_k W_k \mathbf{S}_k$ is a statistically calculated vector photon-distribution-factor, $\mathbf{S}_k$ denotes the polarization state vector of the $k$-th detected photon bundle. For a radiance detector at direction $\mathbf{\Omega}$ with acceptance solid angle of $\Delta\Omega$ and projection area of $A_\perp$, the polarized radiance $\mathbf{I}(\mathbf{\Omega})$ is calculated from

$$\mathbf{I}(\mathbf{\Omega}) = \frac{P_{src}}{A_\perp \Delta\Omega}\mathbf{D}_f \tag{31}$$

## 4. Results and discussion

In order to verify the performance of the MC method presented above and to provide examples for comparison with other models, three types of cases are presented. The first is a case for polarized radiative transfer in a non-participating medium with linear gradient index distribution. Analytical results can be obtained for this case due to its simplicity. The second case considers a Mie scattering medium with collimated beam irradiation. The third case is about sunlight irradiation in a Rayleigh scattering atmosphere with horizontal ray path, which is the path used in limb remote sensing. It will be demonstrated that gradient index distribution will significantly affect both the angular distribution of intensity and the polarization state. The importance of gradient index distribution in atmosphere on polarized light transport is illustrated. The simulation time was evaluated for the simulation of different cases using a computer with a 2.5GHz CPU.

### 4.1. Case 1: non-participating slab with linear refractive index distribution

To verify the presented method, polarized radiative transfer in a non-scattering and non-absorbing slab with gradient index distribution is considered. The refractive index distribution in the slab is given as

$$n(z) = 1.2 + 0.6z / H \tag{32}$$

in which $H$ is the height of the slab. Because the medium is non-participating, the Stokes parameters will follow the generalized Clausius invariant for plane ray path [18], namely,

$$\frac{\mathrm{d}}{\mathrm{d}s} \left[ \frac{\mathbf{I}}{n^2} \right] = 0 \tag{33}$$

Hence there will be $\mathbf{I}(s_1) / n^2(s_1) = \mathbf{I}(s_2) / n^2(s_2)$ for two points on the ray path $s_1$ and $s_2$. The Stokes parameters vector on the boundary will determine the intensity distribution along the whole path. As such, the angular distribution of $\mathbf{I}$ can be obtained analytically. Since all the four component of $\mathbf{I}$ follow the same law in this case, here only the results of radiative intensity are presented.

There are two different conditions of ray path, a beam transfer along z-direction or the reverse. The beam will converge in the former condition and will diverge in the latter for the given refractive index distribution. Here light irradiation both from the top boundary ( $z = H$ ) and from the bottom boundary ( $z = 0$ ) are simulated. Figure 3 shows the simulated angular distributions of radiative intensity at different locations in the slab by the MC method, in which the horizontal axis is $\xi = \cos\theta$, $\theta$ is the incident polar angle. The analytical results are also shown for comparison. In the MC simulations, $1 \times 10^7$ photon bundles were used. Since the mean free path will be infinity for non-participating medium, a small optical thickness is taken as an approximation, e.g., $\tau = \beta H = 0.001$. Intensity detectors with infinite projection area were applied and only the photon bundles fell within a cone with solid angle of $\Delta\Omega = 0.01$ were recorded.

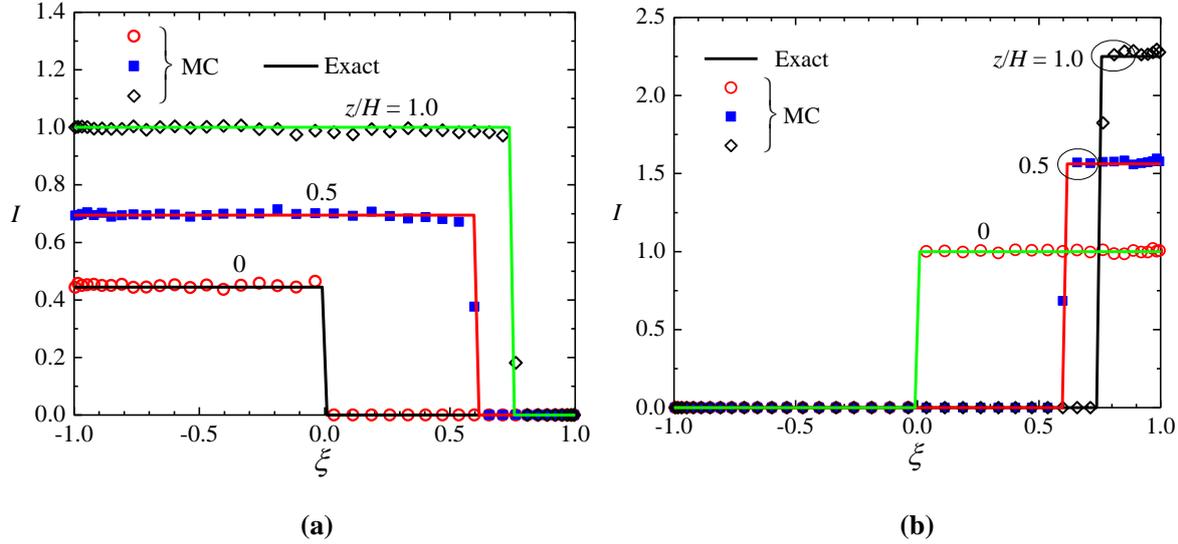

**(a)**             **(b)**

**Fig. 3.** Angular distributions of radiative intensity at different locations in the gradient index slab, **(a)** beam irradiated from the top boundary, **(b)** beam irradiated from the bottom boundary.

The results presented in Fig. 3(a) and 3(b) is for the case with a unit diffuse source ($I(\theta) = 1$)

irradiated from the top boundary and from the bottom boundary, respectively. From Fig. 3(a), it is seen that the radiative intensity decreases when the location approaches the bottom boundary, which is due to the decrease of refractive index as indicated by Eq. (33). It is also observed that there are some angular range in

$\xi > 0$ where the radiative intensity is non-zero for $z/H = 0.5$ and 1.0. Since the diffuse source irradiates

at $\xi < 0$, this means some of the photons transport to the backward direction, which is attributed to total internal reflection. From Fig. 3(b), it is seen that the angular distribution range of radiative intensity converges with increasing $z$. In this case, the diffuse source is irradiated at $z = 0$, the beam will converge with the increase of refractive index and total internal reflection will not occur. Meanwhile, the magnitude of radiative intensity increases. As seen from the figure, the results obtained by the MC method agree well with the analytical results within reasonable statistical errors. The location of discontinuities in the intensity distribution are accurately captured, and the beam converging and diverging effects caused by the gradient refractive index distribution are well predicted.

### 4.2. Case 2: Mie scattering of collimated beam in a gradient-index medium

In this case, the combined effects of absorbing, scattering and gradient index distribution are considered. The case is similar to the 'L=13 problem' of Garcia and Siewert [50], the difference is that the medium is considered to have a refractive index distribution given by Eq. (32). An oblique beam (oblique angle is $\cos\theta_c = 0.2$, and $\varphi_c = \pi/2$) penetrates into a scattering medium as shown in Fig. 4.

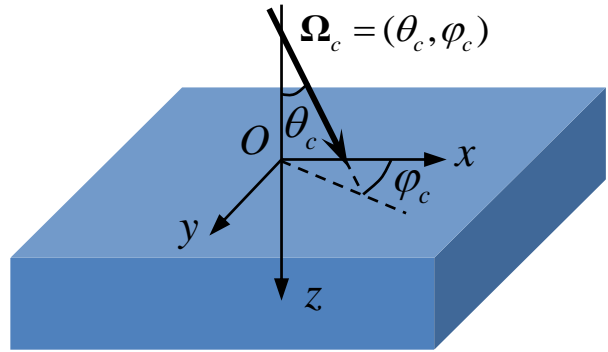

**FIG. 4.** Schematic of the oblique incident collimated beam irradiating into a layer of atmosphere.

The irradiation flux density of the beam is $\pi$. The optical thickness based on the thickness of the medium $H$ is 1.0, and the single scattering albedo is 0.99. The phase matrix is for Mie scattering at a wavelength of 0.951 μm from a gamma distribution of particles with effective radius of 0.2 μm, effective variance of 0.07, and refractive index $n$=1.44. The Legendre coefficients for the phase matrix were used in the scattering calculation, which were obtained by Evens and Stephen [27]. The lower boundary of the atmosphere is diffuse reflection with $\rho = 0.1$. The emission of the atmosphere and the boundary are omitted.

In order for validation, the original '$L$=13 problem', namely, with a uniform refractive index distribution, is simulated first. The MC method is applied to obtain the angular distribution of Stokes parameters at plane $\varphi$ =0. Figure 5 shows the angular distribution of the four Stokes parameters obtained using the MC method at $z/H$ =0.5. The results obtained by Garcia and Siewert [50] using the $F_N$ method and by using our previous developed spectral element method (SEM) [51] are also shown for comparison. In each of the MC simulation, $3 \times 10^8$ photon bundles were launched. The simulation time is about one hour. All the four components of Stokes parameters obtained by the MC results agree well with the reference results, the relative error is less than 1% at most data points.

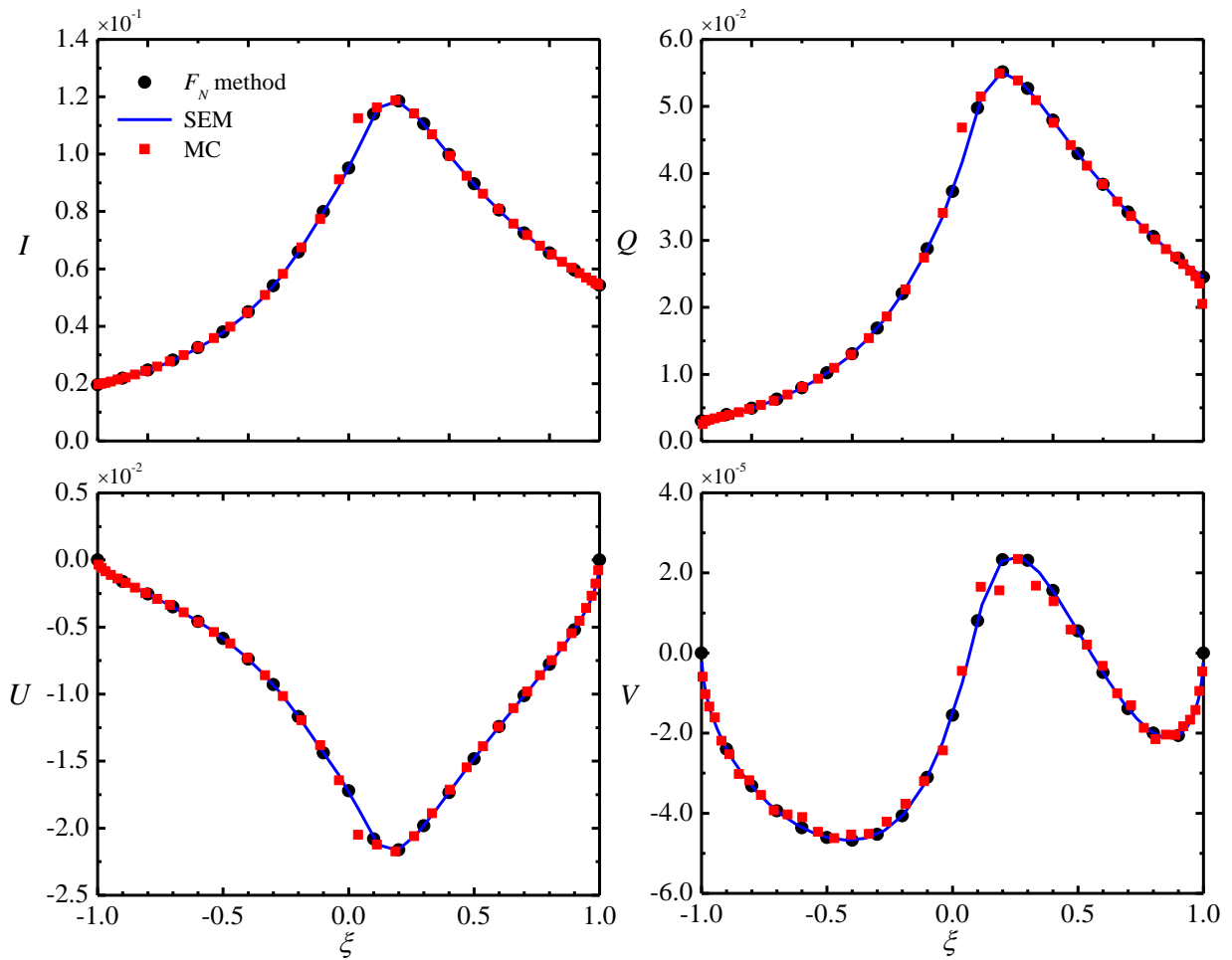

**FIG. 5.** Angular distribution of the four Stokes parameters for the '$L$=13 problem' solved by different methods at location $z/H$ =0.5.

Then the MC method is applied to the case of gradient index distribution. Figure 6(a), (b) and (c) shows the angular distribution of the four Stokes parameters obtained using the MC method at $z/H = 0$, 0.5 and 1, respectively. The Stokes parameters distribution obtained under uniform refractive index distribution solved by the $F_N$ method [50] are also shown for comparison. In each MC simulation for gradient-index media, $5 \times 10^8$ photon bundles were launched. The simulation time is about 6 hour. It is observed that the gradient index distribution greatly affects the angular distribution of Stokes parameters. At $z/H = 0$ shown in Fig. 6(a), the trends of angular distribution of Stokes parameters for uniform and gradient index distribution are similar, however, the magnitude are mostly enhanced except for the $V$ component at $\xi < -0.35$.

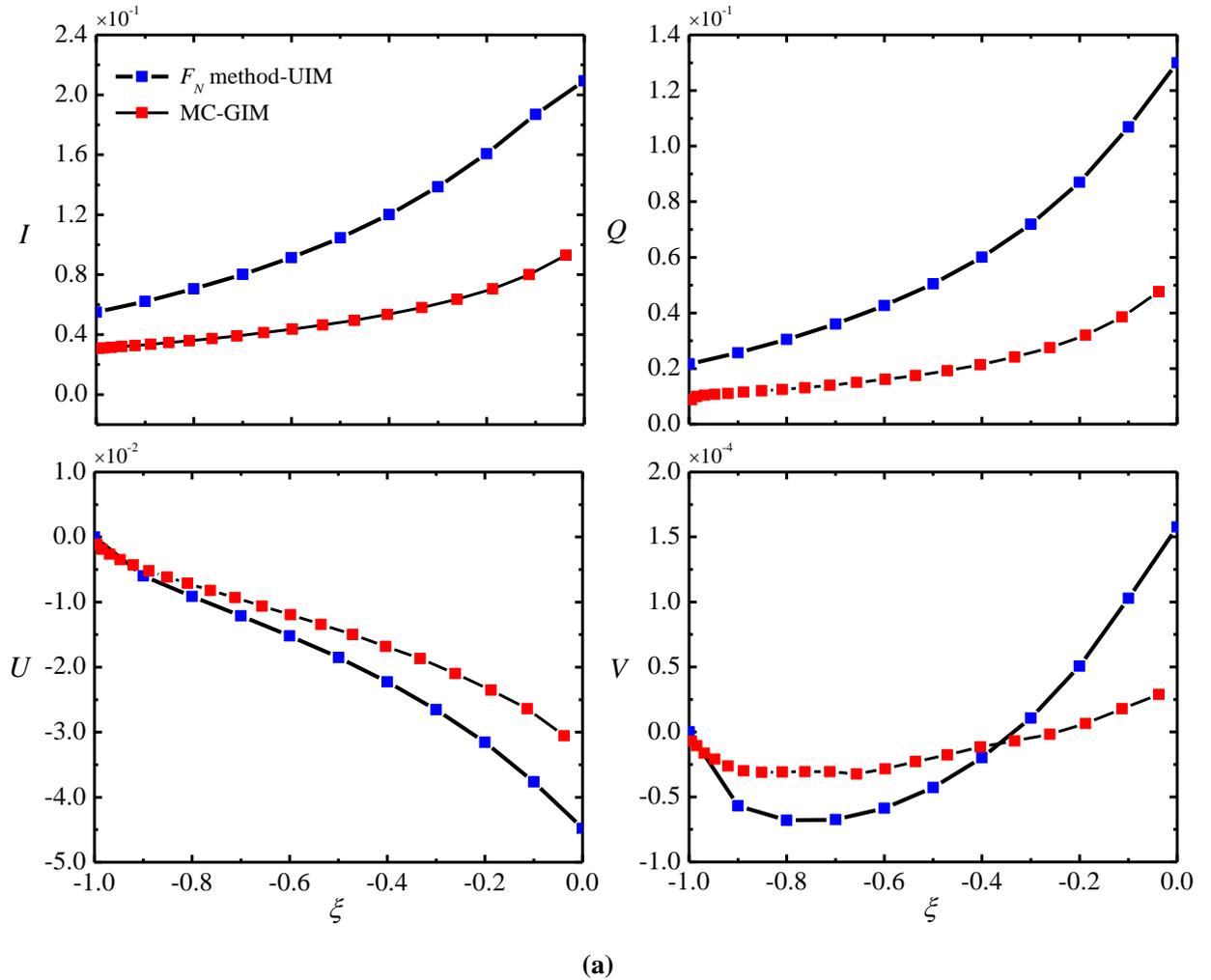

**(a)**

**FIG. 6.** Angular distribution of the four Stokes parameters for the case of gradient refractive index distribution solved by the MC method (MC-GIM) and for the case of uniform index distribution solved by the FN method ($F_N$-UIM) at three different positions, **(a)** $z/H = 0$, **(b)** $z/H = 0.5$ and **(c)** $z/H = 1$.

At the direction close to z-axis ($\xi = -1$), the enhancement is the smallest. This can be understood that the ray geometry is more close to straight line at these directions, hence the effect of gradient index distribution is the smallest. At $z / H = 0.5$ Fig. 6(b), the location of intensity peak at downward direction moves to bigger value of $\xi$, indicating the peak is shifted close to the z-axis direction. Furthermore, the width of the peaks become narrower. Both of these are attributed to the beam converging effect caused by gradient index distribution when the beam transfer along the direction where the refractive index gradually increases.

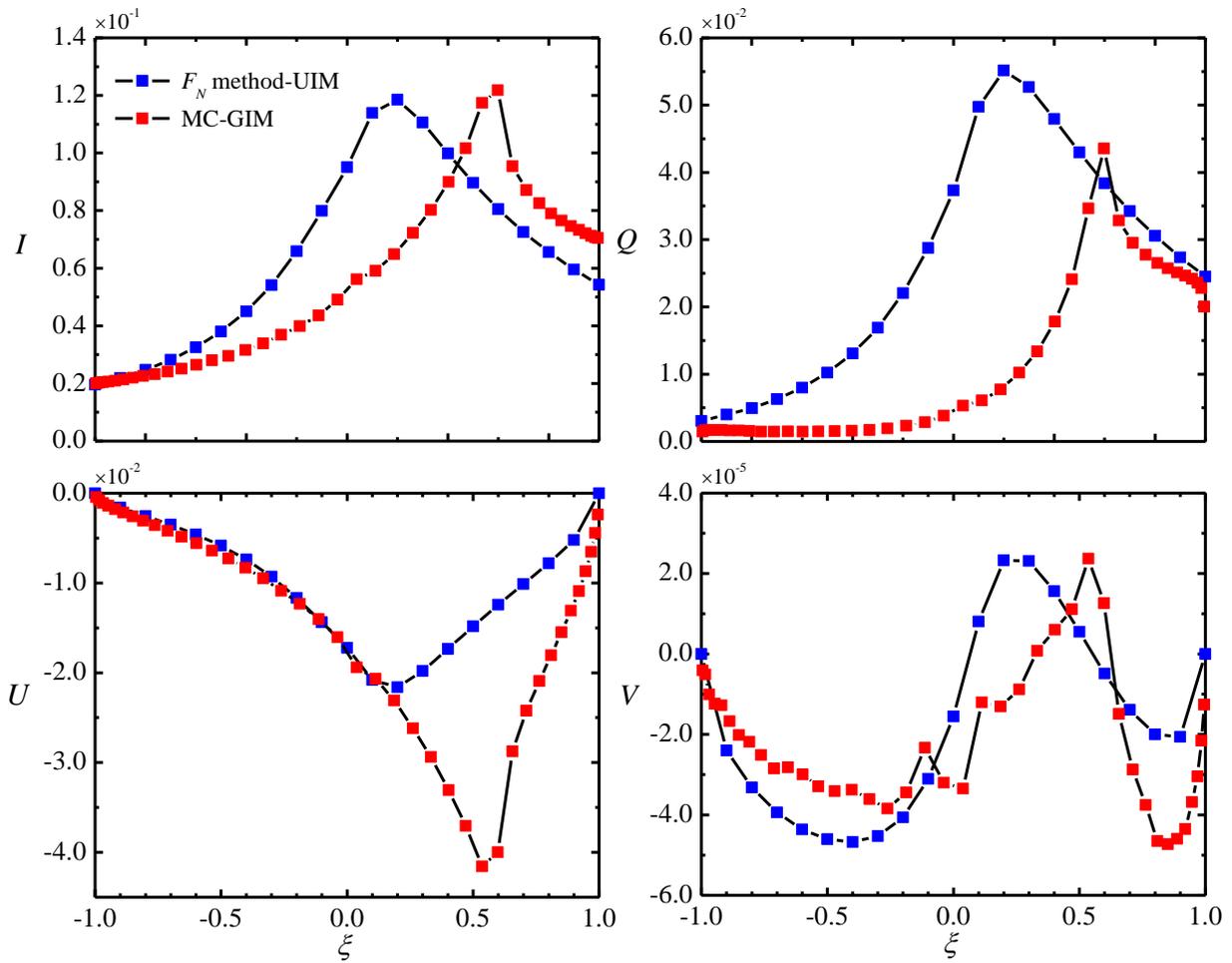

**(b)**

**FIG. 6.** Continued

At $z / H = 1.0$ Fig. 6(c), the effect of gradient index distribution turns to be more complex. The flatten peak in the downward direction under uniform index distribution becomes a narrower peak for the case of gradient index distribution. The intensity distribution in the upward direction ($\xi < 0$) caused by reflection at the bottom boundary is increased, which can be understood that the gradient index distribution induced converging effect enlarges the incident beam intensity. The gradient index distribution almost totally alters the angular distribution of the other three components of Stokes parameters at downward directions. It is difficult to interpret the phenomena since the combined processes of scattering and transfer along curved ray path is very complex. The effect of gradient index distribution on the angular distribution of polarization state in scattering media still needs to be studied further.

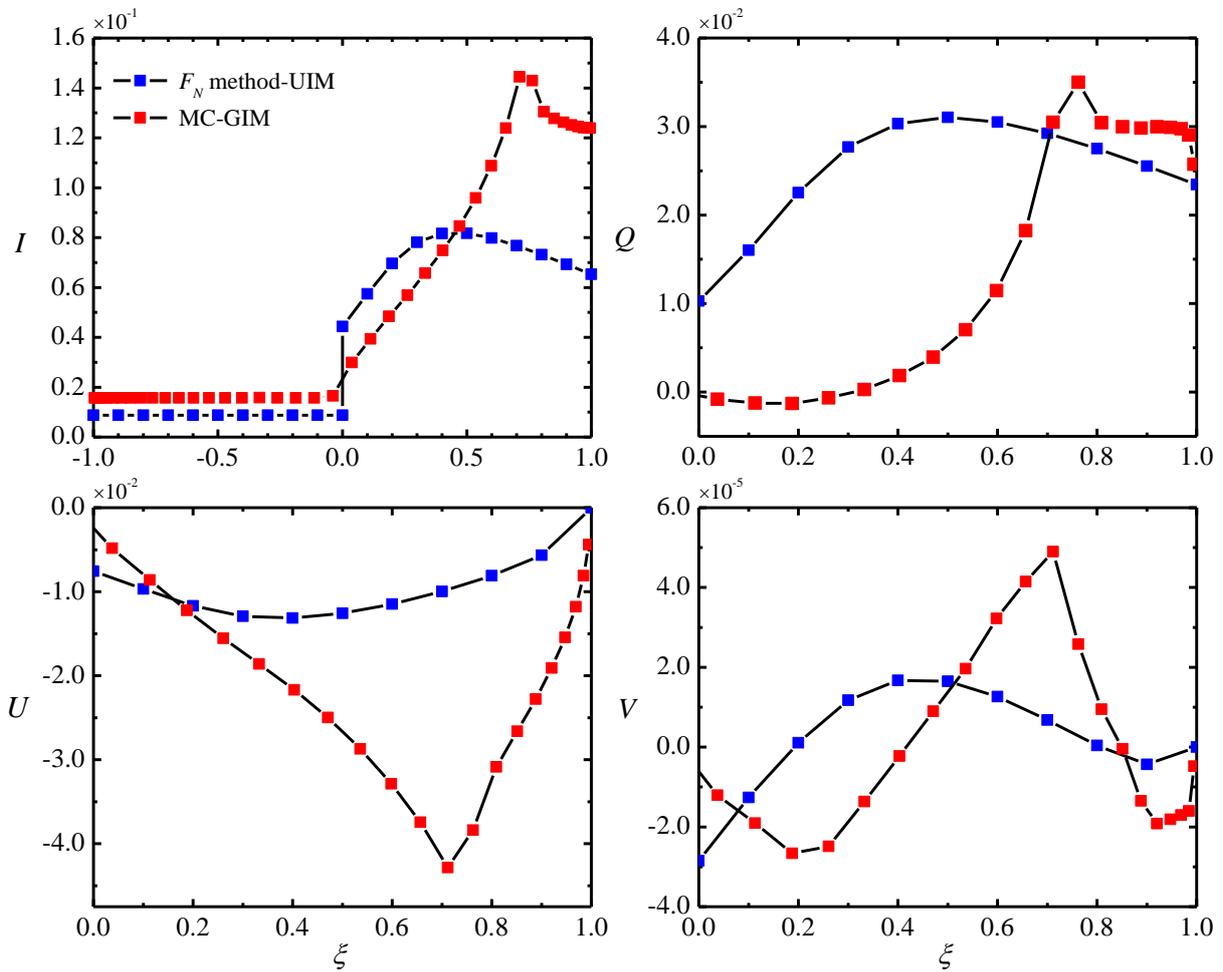

**(c)**

**FIG. 6.** Continued

### 4.3. Case 3: Rayleigh scattering atmosphere with gradient index distribution

In the previous two cases, an assumed refractive index distribution was used for demonstration. In this case, we consider a realistic refractive index distribution in the atmosphere. The 1976 version of U.S.

standard atmosphere model is used here, which is also a widely used atmosphere model in literatures. The standard atmosphere model defines the values for atmospheric temperature, density, pressure and other properties over a wide range of altitudes.

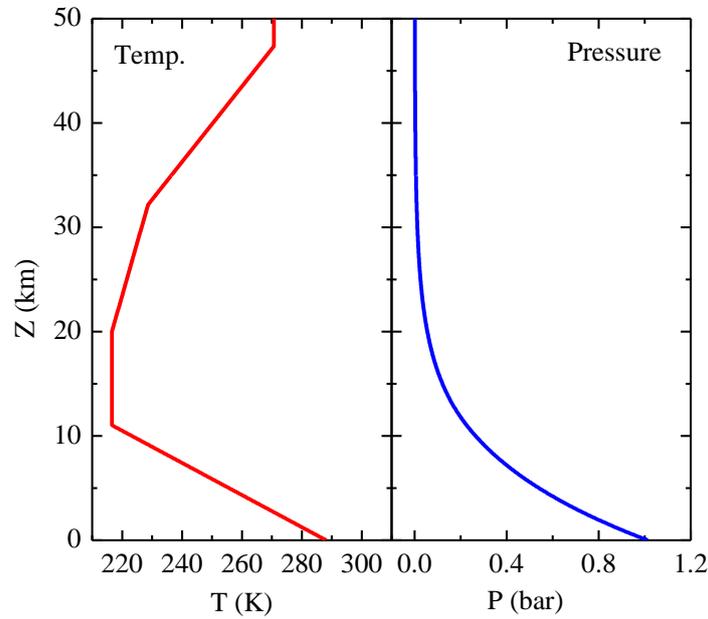

**(a)**

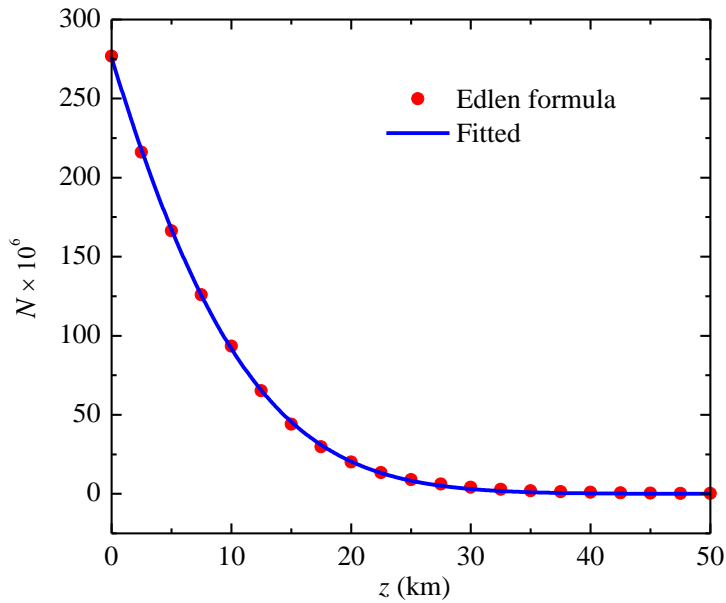

**(b)**

**FIG. 7. (a)** The profiles of temperature and pressure for the standard atmosphere, **(b)** The refractivity profile for the standard atmosphere calculated based on Eq. (34) for a wavelength of 0.6328 μm.

The profiles of temperature and pressure are two key parameters to estimate the refractive index distribution, which are shown in Fig. 7(a) for the standard atmosphere. The refractivity of dry air is calculated based on Edlen's semi-empirical law [6, 9, 52] as

$$N \times 10^6 = \left( 776.2 + 4.36 \times 10^{-8} \nu^2 \right) \frac{P}{T} \tag{34}$$

where $N = n - 1$ is the refractivity, $\nu$ is wave number (cm$^{-1}$), $P$ is the pressure of dry air (kPa) and $T$ is the air temperature (K). This model is considered to be valid for dry air at altitudes less than 100 km where the mixing ratio between oxygen and nitrogen is fixed and for wavelength from 0.2 to 2,000 μm. The refractivity profile for the standard atmosphere calculated based on Eq. (34) for a wavelength of 0.6328 μm is plotted in Fig. 7(b). A fitted formula for the refractivity is obtained (also shown in Fig. 7(b)) to ease the curved ray tracing in the MC simulation, which is in the form of exponential function and given as follows.

$$N_{\text{fit}} \times 10^6 = \exp \left[ c_1 + c_2 z + c_3 z^2 \right] \tag{35}$$

where $z$ is the altitude in kilometer, $c_1 = 5.62$, $c_2 = -0.09$, and $c_3 = -0.002$. The refractive index is obtained as $n = 1 + N_{\text{fit}}$, and the gradient of refractive index distribution can be obtained analytically by using the fitted formula.

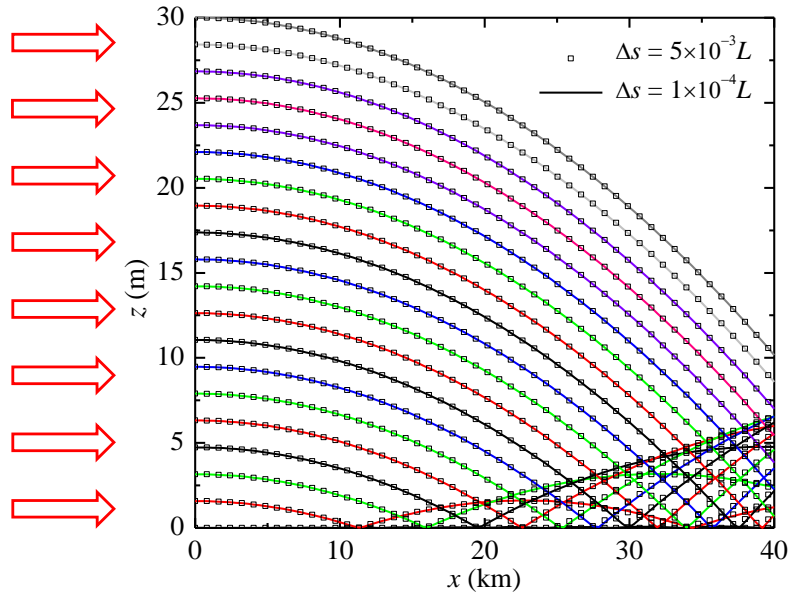

**FIG. 8.** The geometry of the problem studied in Case 3. A collimated beam is irradiated from left. Also shown are the ray paths traced using step size $\Delta s = 5 \times 10^{-3} L$ and $1 \times 10^{-4} L$, where $L$ is the width of the domain.

The geometry of the problem is shown in Fig. 8, a collimated light beam with unit radiative flux density is irradiated horizontally to atmosphere from left to right (in the $x$-direction). The height and width of the simulated region is $H = 30$ m and $L = 40$ km, respectively, and the dimension in perpendicular to the paper ($y$-direction) is assumed infinite. The curvature of the earth surface is neglected. Rayleigh scattering is considered with a single scattering albedo $\omega = 0.99$. The ground is assumed to be water with a refractive index 1.33. The radiative flux density of Stokes parameters along the boundaries are recorded. All the photon bundles moving out the simulation domain are terminated. A raytracing test was performed first in order for choosing an appropriate tracing step size. Figure 8 shows the ray paths traced using step size $\Delta s = 5 \times 10^{-3} L$ and $1 \times 10^{-4} L$. The test demonstrates that $\Delta s = 5 \times 10^{-3} L$ is sufficient to give accurate ray paths and used in the following simulations. Note that the rays bend toward the ground due to greater refractive index near the ground and reflected at the water surface.

As for verification, we consider a case that the width of the simulation domain is only $L = 30$ m, in this case, the gradient of refractive index can be neglected due to the very short distance of ray path traced in the simulation domain. Hence the recorded polarized radiative flux density can be compared with the results obtained using polarized radiative transfer code for uniform refractive index media. The flux density distribution were recorded by two-dimensional detectors with detecting area of $A = 1$m. Figure 9(a) and (b) show the dimensionless polarized radiative flux density distribution along the right and top boundary, respectively, obtained using the MC method. The irradiated collimated beam has unit flux density and a polarization state vector $(1, 0.5, 0.5, 0.5)$. The optical thickness based on the width of the simulation domain is 1.0. The results obtained using SEM [51] are also shown for reference. Along the right boundary, each of the four components of polarized radiative flux density has a peak at the middle location. Along the top boundary, only $I$ and $Q$ components polarized radiative flux density have a peak located close to the light source. The results of MC are shown to agree very well with the reference results.

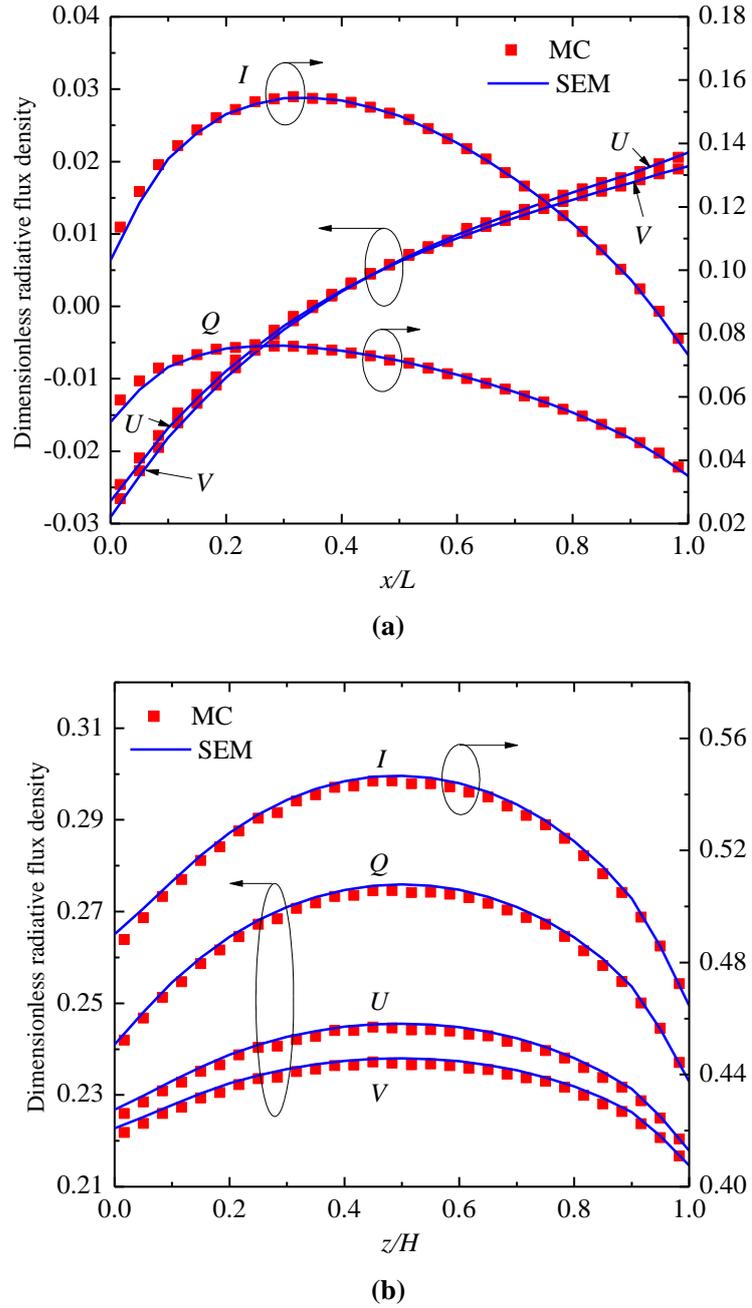

**FIG. 9.** The dimensionless polarized radiative flux density distribution along different boundaries, **(a)** right, **(b)** top. The width of the simulation domain is $L = 30$ m and the irradiated collimated beam has unit flux density and a polarization state vector of $(1, 0.5, 0.5, 0.5)$.

Now we consider the case with width $L = 40$km. When a ray transfers from the source to the right boundary, it will travel a much long distance. In this case, the gradient index distribution of atmosphere will have significant effect on radiative transfer as indicated by the ray path shown in Fig. 8. Figure 10 presents the simulated dimensionless polarized radiative flux density distribution along the right boundary at three atmosphere optical thickness based on $L$, namely, $\tau_L$ =0.1, 0.5 and 1.0. The irradiated collimated beam is unpolarized with unit flux density.

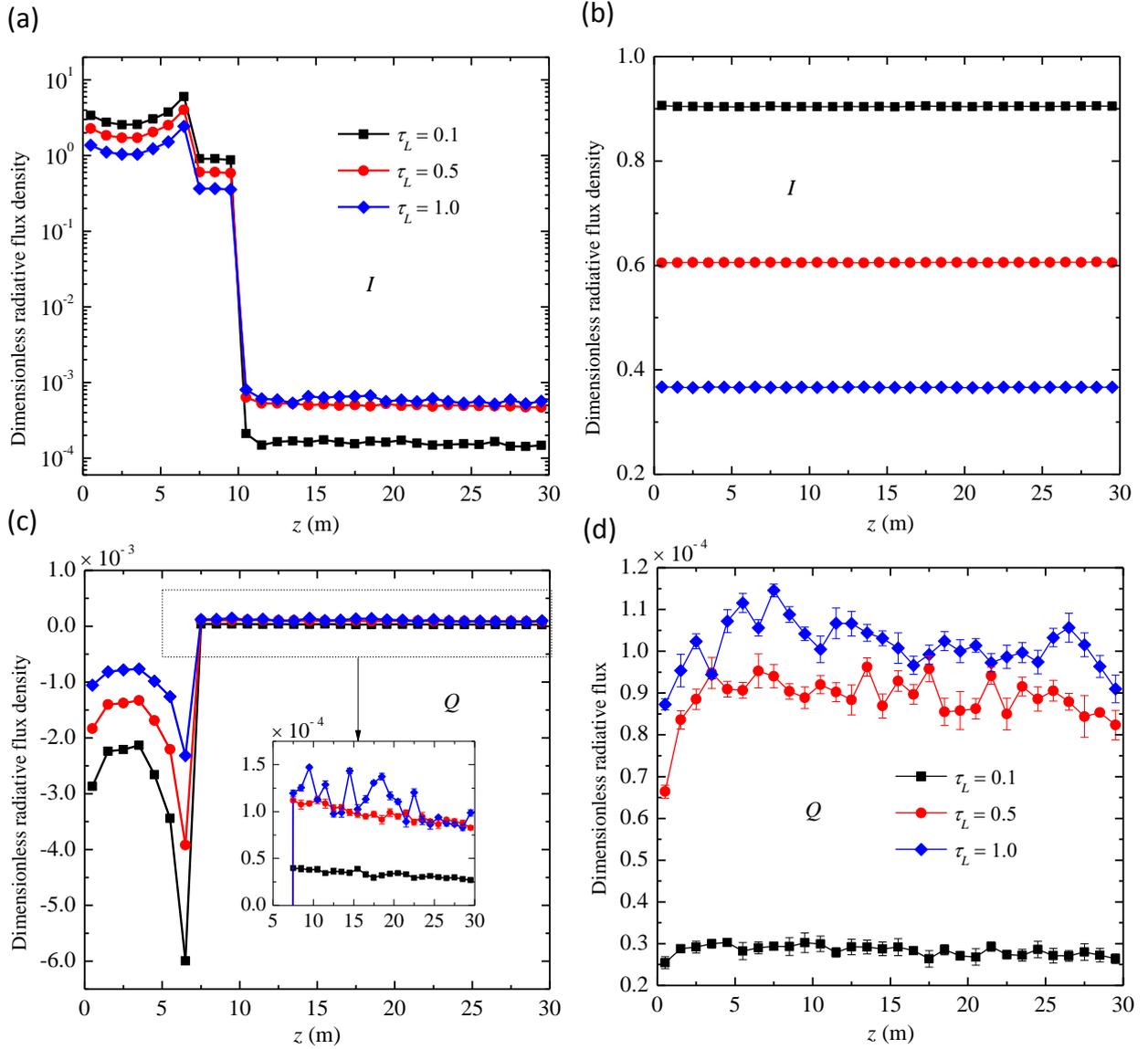

**FIG. 10.** The dimensionless polarized radiative flux density distribution along the right boundary at three atmosphere optical thickness based on $L$, namely, $\tau_L$ = 0.1, 0.5 and 1.0, **(a)** and **(c)** are the $I$ and $Q$ components for the gradient index atmosphere, **(b)** and **(d)** are the $I$ and $Q$ components for the uniform index atmosphere. The inset gives an enlarged view of the $Q$ component flux density distribution. The error bars show the standard deviation of the Monte Carlo simulated results.

Figure 10 (a) and (c) gives the results of $I$ and $Q$ components, respectively. Figure 10 (b) and (d) gives the corresponding results obtained by assuming uniform index distribution for comparison. In the MC simulations, $1 \times 10^8$ photon bundles were launched. The error bars in Fig. 10 show the standard deviation of the MC simulated results, which were obtained based on 5 MC simulations. The standard deviations are on the order of $1 \times 10^{-6}$, which is too small and not shown for the plot of radiative intensity. The simulation time required for the case $\tau_L = 0.1$ was about 3 hours, and about 8 hours for $\tau_L = 1.0$. It is observed that there exists a height range, i.e. $z < 10$m, where the radiative flux density has a high value (as shown in Fig. 10(a)), which is even higher than 3 times of the radiative flux density of the source, e.g. for $\tau_L = 0.1$. This is attributed to the ray bending toward the ground as shown in Fig. 8. Furthermore, due to the reflection at water surface, a sharp enhancement of radiative flux density is induced close to $z = 7$m. At $z > 10$m, nearly no direct photons can reach the detectors (shown in Fig. 8). Hence the detected photon flux is from the scattered photons. As shown in Fig. 10(a), the dimensionless radiative flux at this height range increases with optical thickness, this is due to the enhanced atmosphere scattering at big value of optical thickness. As for the uniform index case shown in Fig. 10(b), the radiative flux density distributions are nearly constant at different height location for different optical thickness and no enhancement are observed. For the $Q$ component radiative flux density shown in Fig. 10(c), similar enhancement is observed at about $z < 7.5$m, where its magnitude has a big value. The magnitude increases with decrease of optical thickness. The $Q$ component is an indication of linear polarization. It has a negative value, which indicates s-polarization. This is an indication of the polarization induced by water surface reflection. At $z > 7.5$m, $Q$ component radiative flux density is positive and has a relative weak magnitude (on the order $10^{-5}$ to $10^{-4}$) as compared to that of $z < 7.5$m, the latter is negative and has magnitude (on the order $10^{-4}$ to $10^{-3}$) about an order greater than the former. Furthermore, as shown in the inset, the magnitude of $Q$ component radiative flux density increases with the increase of optical thickness. This is an indication of scattering enhancement. As for the uniform index case shown in Fig. 10(d), the $Q$ component radiative flux density is positive at different height location for different optical thickness, and the magnitude is on the order $10^{-5}$ to $10^{-4}$. With the increase of optical thickness, the magnitude of $Q$ component radiative flux density increases, indicating the polarization induced by scattering process. By comparison, it is shown that the gradient index distribution in atmosphere will have significant effect on polarized light transport, especially for long distance transfer. It can enhance both the detected radiative flux density and the degree of polarization under certain configurations, e.g., the case studied here.

## 5. Conclusions

A Monte Carlo method is presented to simulate polarized radiative transfer in gradient-index media, which is demonstrated to be effective to solve polarized radiative transfer problems. The gradient index distribution will make light beam converge or diverge during transfer and induce rotation of polarization ellipse. Furthermore, totally internal reflection may happen for some incident direction. These effects significantly influence the angular distribution of Stokes parameters, even totally alter the angular distribution. As such, omitting gradient index distribution should be taken very carefully.

The combined process of scattering and transfer along curved ray path is very complex, which makes it difficult to interpret the results. For atmosphere with realistic refractive index distribution, it is shown that the gradient index distribution plays an important role for long distance polarized light transport. The gradient index distribution can enhance both the detected radiative flux density and the degree of polarization. Further study on the effect of gradient index distribution on polarized radiative transfer in scattering media, developing more efficient and accurate solution methods, and application of the theory to engineering problems are very appealing.

## Acknowledgments

J.M.Z and J.Y. T. thanks the support by the National Nature Science Foundation of China (Nos.

51076038, 51276051) and the Fundamental Research Funds for the Central Universities (Grant no. HIT. NSRIF. 2013094). L.H.L. thanks the support by the National Nature Science Foundation of China (No. 51121004).

**References**

[1]  Tsang L, Kong JA and Shin RT. Theory of microwave remote sensing. New York: Wiley; 1985.

[2]  Goloub P, Herman M, Chepfer H, Riedi J, Brogniez G, Couvert P and Seze G. Cloud thermodynamical phase classification from the POLDER spaceborne instrument. J Geophys Res 2000 105: 14747- 59.

[3]  Deiveegan M, Balaji C and Venkateshan SP. A polarized microwave radiative transfer model for passive remote sensing. Atmospheric Research 2008; 88: 277-93.

[4]  Noerdlinger PD. Atmospheric refraction effects in earth remote sensing. ISPRS Journal of Photogrammetry and Remote Sensing 1999; 54: 360-73.

[5]  van der Werf SY, Konnen GP and Lehn WH. Novaya zemlya effect and sunsets. Appl Opt 2003; 42: 367-78.

[6]  Thomas ME and Joseph RI. Astronomical refraction. Johns Hopkins APL Technical Digest 1996; 17: 279.

[7]  Platt U, Pfeilsticker K and Vollmer M. Radiation and optics in the atmosphere. In: Träger F, editors. Springer handbook of lasers and optics, vol. New York: Springer; 2012. p. 1475-517.

[8]  Weichel H. Laser beam propagation in the atmosphere. Washington: SPIE press; 1990.

[9]  Neda Z and Volkan-Kacso S. Flatness of the setting sun. American Journal of Physics 2003; 71: 379-85.

[10] Uchiyama H, Nakajima M and Yuta S. Measurement of flame temperature distribution by IR emission computed tomography. Appl Opt 1985; 24: 4111-6.

[11] Liu LH and Man GL. Reconstruction of time-averaged temperature of non-axisymmetric turbulent unconfined sooting flame by inverse radiation analysis. JQSRT 2003; 78: 139-49.

[12] Zhou H-C, Lou C, Cheng Q, Jiang Z, He J, Huang B, Pei Z and Lu C. Experimental investigations on visualization of three-dimensional temperature distributions in a large-scale pulverized-coal-fired boiler furnace. Proceedings of the Combustion Institute 2005; 30: 1699-706.

[13] Stavroudis ON. The optics of rays, wavefronts and caustics. New York: Academic; 1972.

[14] Karavstov YA and Orlov YI. Geometrical optics of inhomogeneous media. New York: Springer-Verlag;

1990.

[15] Qiao YT. Graded index optics. Beijing: Science Press; 1991.

[16] Chandrasekhar S. Radiative transfer. NewYork: Oxford University Press; 1950.

[17] Lau CW and Watson KM. Radiation transport along curved ray paths. J Math Phys 1970; 11: 3125-37.

[18] Zhao JM, Tan JY and Liu LH. On the derivation of vector radiative transfer equation for polarized radiative transport in graded index media. JQSRT 2012; 113: 239-50.

[19] Vaillon R, Wong BT and Menguc MP. Polarized radiative transfer in a particle-laden semi-transparent medium via a vector Monte Carlo method. JQSRT 2004; 84: 383-94.

[20] Ramella-Roman JC, Prahl SA and Jacques SL. Three Monte Carlo programs of polarized light transport into scattering media: Part I. Optics Express 2005; 13: 4420-38.

[21] Tynes HH, Kattawar GW, Zege EP, Katsev IL, Prikhach AS and Chaikovskaya LI. Monte Carlo and multicomponent approximation methods for vector radiative transfer by use of effective mueller matrix calculations. Appl Opt 2001; 40: 400-12.

[22] Davis C, Emde C and Harwood R. A 3D polarized reversed Monte Carlo radiative transfer model for mm and sub-mm passive remote sensing in cloudy atmospheres. IEEE Trans Geosci Remote Sens 2005 43: 1096-101.

[23] Raković MJ, Kattawar GW, Mehruűbeoğlu M, Cameron BD, Wang LV, Rastegar S and Coté GL. Light backscattering polarization patterns from turbid media: Theory and experiment. Appl Opt 1999; 38: 3399-408.

[24] Weng F. A multi-layer discrete-ordinate method for vector radiative transfer in a vertically-inhomogeneous, emitting and scattering atmosphere-I. Theory. JQSRT 1992; 47: 19-33.

[25] Siewert CE. A discrete-ordinates solution for radiative-transfer models that include polarization effects. JQSRT 2000; 64: 227- 54.

[26] Haferman JL, Smith TF and Krajewski WF. A multi-dimensional discrete-ordinates method for polarized radiative transfer. Part I: Validation for randomly oriented axisymmetric particles. JQSRT 1997; 58: 379-98.

[27] Evans KF and Stephens GL. A new polarized atmospheric radiative transfer model. JQSRT 1991; 46: 413~23.

[28] Ben Abdallah P and Le Dez V. Thermal emission of a semi-transparent slab with variable spatial refractive index. JQSRT 2000; 67: 185-98.

[29] Huang Y, Xia XL and Tan HP. Temperature field of radiative equilibrium in a semitransparent slab with a

linear refractive index and gray walls. JQSRT 2002; 74: 249-61.

[30]Liu LH. Discrete curved ray-tracing method for radiative transfer in an absorbing-emitting semitransparent slab with variable spatial refractive index. JQSRT 2004; 83: 223-8.

[31]Liu LH, Zhang HC and Tan HP. Monte Carlo discrete curved ray-tracing method for radiative transfer in an absorbing-emitting semitransparent slab with variable spatial refractive index. JQSRT 2004; 84: 357-62.

[32]Wu C-Y. Monte Carlo simulation of transient radiative transfer in a medium with a variable refractive index. Int J Heat Mass Transfer 2009; 52: 4151-9.

[33]Liu LH. Finite volume method for radiation heat transfer in graded index medium. J Thermophys Heat Transfer 2006; 20: 59-66.

[34]Asllanaj F and Fumeron S. Modified finite volume method applied to radiative transfer in 2D complex geometries and graded index media. JQSRT 2010; 111: 274-9.

[35]Zhang L, Zhao JM, Liu LH and Wang SY. Hybrid finite volume/ finite element method for radiative heat transfer in graded index media. JQSRT 2012; 113: 1826-35.

[36]Liu LH, Zhang L and Tan HP. Finite element method for radiation heat transfer in multi-dimensional graded index medium. JQSRT 2006; 97: 436-45.

[37]Liu LH. Finite element solution of radiative transfer across a slab with variable spatial refractive index. Int J Heat Mass Transfer 2005; 48: 2260-5.

[38]Zhao JM and Liu LH. Solution of radiative heat transfer in graded index media by least square spectral element method. Int J Heat Mass Transfer 2007; 50: 2634-42.

[39]Wang Z, Cheng Q, Wang G and Zhou H. The dresor method for radiative heat transfer in a one-dimensional medium with variable refractive index. JQSRT 2011; 112: 2835-45.

[40]Garcia RDM. Fresnel boundary and interface conditions for polarized radiative transfer in a multilayer medium. JQSRT 2012; 113: 306-17.

[41]Garcia RDM. Radiative transfer with polarization in a multi-layer medium subject to Fresnel boundary and interface conditions. JQSRT 2013; 115: 28-45.

[42]Ben X, Yi HL and Tan HP. Approximate solution to vector radiative transfer in gradient-index medium. Appl Opt 2014; 53: 388-401.

[43]Born M and Wolf E. Principles of optics. 7th ed. Cambridge: Cambridge University Press; 1970.

[44]Pomraning GC. The equations of radiation hydrodynamics. New York: Pergamon; 1973.

[45]Goldstein DH. Polarized light. 2nd ed. New York: Marcel Dekker, Inc.; 2003.

[46]Bohren CF and Huffman DR. Absorption and scattering of light by small particles. New York: Wiley; 1998.

[47]van de Hulst HC. Light scattering by small particles. New York: Dover; 1981.

[48]Modest MF. Radiative heat transfer. 2nd ed. New York: Academic Press; 2003.

[49]Zhai P-W, Hu Y, Chowdhary J, Trepte CR, Lucker PL and Josset DB. A vector radiative transfer model for coupled atmosphere and ocean systems with a rough interface. JQSRT 2010; 111: 1025-40.

[50]Garcia RDM and Siewert CE. The FN method for radiative transfer models that include polarization effects. JQSRT 1989; 41: 117-45.

[51]Zhao JM, Liu LH and Hsu p-f. Spectral element method for vector radiative transfer equation. JQSRT 2009; 111: 433-46.

[52]Edlen B. The refractive index of air. Metrologia 1967; 2: 71-80.